\documentclass[showpacs,eqsecnum,preprint]{revtex4}
\usepackage{amsmath,amssymb} 

\begin{document}

\title{Non-linear $\sigma$ models solvable by the
Aratyn-Ferreira-Zimerman Ansatz}

\author{Minoru Hirayama and Chang-Guang Shi}

\affiliation{Department of Physics, Toyama University,
 Toyama 930-8555,Japan}

\begin{abstract}

Nonlinear $\sigma$ models compatible with the aratyn-Ferreira-Zimerman ansatz
are discussed, the latter ansatz automatically leading to configurations
with definite values of the Hopf index. These models are allowed to involve a weight
factor which is a function of one of the toroidal coordinates. Depending on
the choice of the weight factor, the field equation takes various forms. In
one model with a special weight factor, the field equation turns out to be
the fifth Painlev\'e equation. This model suggests the existence of a knot
soliton strictly confined in a finite spatial volume. Some other
interesting cases are also discussed.

\end{abstract}


\maketitle

\section{Introduction}

In some recent papers, it was suggested that a class of relativistic
field theories might possess soliton solutions of knot type. Among
them, the Faddeev-Skyrme $O(3)$ nonlinear $\sigma$ model has been
investigated most intensively\cite{L.FaddeevAJ,FaddeevAJ,Gladikowski,Hietarinta,Battye}. This model concerns the real scalar
fields 
\begin{equation}
\boldsymbol{n}(x)=(n^1(x),n^2(x),n^3(x))
\end{equation}
 satisfying
\begin{equation}
{\boldsymbol{n}}^2(x)=\boldsymbol{n}(x)\cdot\boldsymbol{n}(x)=\sum\limits_{a=1}^{3}n^a(x) n^a(x)=1. 
\end{equation}
The Lagrangian density of this model is given by
\begin{eqnarray}
{\mathcal L}_F(x) &=&c_2 l_2(x)+c_4 l_4(x),\\
l_2(x)&=&\partial_{\mu}\boldsymbol{n}(x)\cdot
 \partial^{\mu}\boldsymbol{n}(x),\\
l_4(x)&=&-H_{\mu\nu}(x) H^{\mu\nu}(x),\\
H_{\mu\nu}(x)&=&\boldsymbol{n}(x)\cdot[\partial_{\mu}\boldsymbol{n}(x)\times
 \partial_{\nu}\boldsymbol{n}(x)]\nonumber\\
&=&\epsilon_{abc}n^a(x)\partial_{\mu}n^b(x)\partial_{\nu}n^c(x), 
\end{eqnarray}
where $c_2$ and $c_4$ are constants. The static energy
functional $E_F[\boldsymbol{n}]$ associated with ${\mathcal L}_F(x)$ is
given by
\begin{eqnarray}
E_F[\boldsymbol{n}]&=&\int dV [c_2 \epsilon_2(\boldsymbol{x})+c_4
 \epsilon_4(\boldsymbol{x})],\\
\epsilon_2(\boldsymbol{x})&=&\sum\limits_{a=1}^{3}\sum\limits_{i=1}^{3}[\partial_i
 n^a(\boldsymbol{x})]^2,\\
\epsilon_4(\boldsymbol{x})&=&\sum\limits_{i,j=1}^{3}[H_{ij}(\boldsymbol{x})]^2,\\
\boldsymbol{x}&=&(x_1,x_2,x_3),\nonumber\\
dV&=&dxdydz
\end{eqnarray}
with $(x_1,x_2,x_3)=(x,y,z)$ being Cartesian coordinates. If we assume
that $\boldsymbol{n}(\boldsymbol x)$ satisfy the boundary condition
\begin{equation}
\boldsymbol{n}(\boldsymbol x)=(0,0,1) \hspace{2mm}\textrm{at} \hspace{2mm}  \arrowvert\boldsymbol{x} \arrowvert=\infty
\end{equation}
and that $\boldsymbol{n}(\boldsymbol x)$ are regular for $ |\boldsymbol{x}|<\infty$, we can regard
$\boldsymbol{n}$ as a mapping from $S^3$ to $S^2$. It is known that such
mappings are classified by the topological number $Q_H[\boldsymbol{n}]$
called the Hopf index, which is also a functional of
$\boldsymbol{n}$.  Vakulenko and Kapitanskii and Kundu and Rybakov found an inequality of the
following type \cite{Vakulenko,Kundu,Ward}:
\begin{equation}
E_F[\boldsymbol{n}]\geqq K\sqrt{c_2 c_4}\biggl
 \arrowvert Q_H[\boldsymbol{n}]\biggr \arrowvert^{3/4},
\end{equation}
where $K$ is equal to $2^{9/2}3^{3/8}\pi^2$. Then a
configuration of  $\boldsymbol{n}(\boldsymbol x)$ with  $Q_H[\boldsymbol{n}]\neq 0$
would be stable against collapsing into a trivial configuration. Through
a numerical analysis aided by a parallel supercomputer, it was revealed
that there indeed exist many interesting solitonic configurations of
$\boldsymbol{n}(\boldsymbol x)$. For example, the configuration with
$Q_H[\boldsymbol{n}]=7$ exhibits the topology of a trefoil knot \cite{Battye}. As for
the model defined by ${\mathcal L}_{\textrm{F}}(x)$, Faddeev and Niemi discussed that
it is intimately related to the low energy dynamics of the $SU(2)$
non-Abelian gauge field \cite{FaddeevJ.}. The generalization to the case of the $SU(N)$
gauge group was also given by them \cite{Faddeev and A.}. 

On the other hand, Aratyn, Ferreira, and Zimerman(AFZ) \cite{AFZ} considered the model
defined by
\begin{equation}   
{\mathcal L}_{AFZ}(x)=[l_4(x)]^{3/4}.
\end{equation}
The static energy functional $E_{\rm AFZ}[\boldsymbol{n}]$ deduced from
${\mathcal L}_{\rm AFZ}(x)$ is invariant under spatial scale transformations,
suggesting the possibility of stable solitonic solutions. With the aid
of an ansatz for $\boldsymbol{n}(\boldsymbol x)$, which we call the AFZ ansatz and
explain in the next section, they reduced the static field equation for
$\boldsymbol{n}(\boldsymbol x)$ to an ordinary differential equation for a single
function and succeeded in solving it. In this way, they obtained an
infinite number of solutions with general values of the Hopf index. To
state their ansatz, they made use of the toroidal coordinates
$(\eta,\xi,\phi)$ related to the Cartesian coordinates $(x,y,z)$ by 
\begin{eqnarray} 
x&=&\frac{a}{q}\hspace{1mm}\textrm{sinh}\hspace{1mm}\eta
 \hspace{1mm}\textrm{cos}\phi,\nonumber\\
y&=&\frac{a}{q}\hspace{1mm}\textrm{sinh}\hspace{1mm}\eta \hspace{1mm}\textrm{sin}\phi,\nonumber\\
z&=&\frac{a}{q}\hspace{1mm}\textrm{sin}\xi,
\end{eqnarray}
where $a$ is a scale parameter of the dimension of length and $q$ is
defined by
\begin{equation} 
q=\textrm{cosh}\eta -\textrm{cos}\xi. 
\end{equation}

In this paper, we consider nonlinear $\sigma$ models defined by the
following ${\mathcal L}_2(x)$, ${\mathcal L}_4(x)$, and ${\mathcal L}_{2+4}(x)$:
\begin{eqnarray}
{\mathcal L}_2(x)&=&\frac{q}{a}\hspace{1mm}\mu(\eta)\hspace{1mm}l_2(x),\\
{\mathcal L}_4(x)&=&\frac{a}{q}\hspace{1mm}\nu(\eta)\hspace{1mm}l_4(x),\\
{\mathcal L}_{2+4}(x)&=&g_2{\mathcal L}_2(x)+g_4{\mathcal L}_4(x),
\end{eqnarray}
where $\mu(\eta)$ and $\nu(\eta)$ are real functions of $\eta$ and $g_2$
and $g_4$ are constants. In contrast to ${\mathcal L}_{AFZ}(x)$, the
time derivatives of fields appear only quadratically in  ${\mathcal
L}_2(x)$, ${\mathcal L}_4(x)$, and ${\mathcal L}_{2+4}(x)$. Just
like $E_{\textrm{AFZ}}[\boldsymbol n]$, the static energy functionals
associated with these models are invariant under spatial scale
transformations. It turns out that the AFZ ansatz is also applicable to
these models. In the case of ${\mathcal L}_4(x)$, the field equation is
solvable for  general $\nu(\eta)$. Thus we also obtain an infinite
number of solitonic solutions with general values of the Hopf index. As for
the case of ${\mathcal L}_2(x)$ with general $\mu(\eta)$, we are led to
a field equation that somewhat resembles the fifth Painlev\'e
equation \cite{Ince,Iwasaki}. For a special choice of $\mu(\eta)$, we indeed obtain the
fifth Painlev\'e equation.  Since the solutions of Painlev\'e equations
have movable poles, we are led to the interesting possibility of knot
solitons strictly confined in finite spatial volumes. For another choice
of $\mu(\eta)$, the field equation is solvable in terms of the Jacobi
elliptic function. Even the complicated case ${\mathcal L}_{2+4}(x)$ can
be discussed through the AFZ ansatz. Although the Faddeev-Skyrme model
is not easy to investigate analytically, the analysis of the models
defined by Eqs. (1.16)-(1.18) seems much easier. Although they are not
Lorentz invariant, we investigate them with the hope that they might
reveal some unexpected features of knot solitons.

This paper is organized as follows. In Sec. II, we review the AFZ ansatz
and the Hopf index briefly. In Sec. III, we discuss the reason why
the forms of ${\mathcal L}_2(x)$ and ${\mathcal L}_4(x)$ are
selected. Models defined by ${\mathcal L}_2(x)$ with various choices of
$\mu(\eta)$ are investigated in  Sec. IV. The field equation associated with
${\mathcal L}_4(x)$ is solved in  Sec. V. In  Sec. VI, we discuss the case where the
Lagrangian density is given by the linear combination ${\mathcal
L}_{2+4}(x)$ of ${\mathcal
L}_2(x)$ and ${\mathcal L}_4(x)$. The final section is devoted to a summary.

\section{Aratyn-Ferreira-Zimerman Ansatz and Hopf index}

To describe the AFZ ansatz, the toroidal coordinates $(\eta,\xi,\phi)$
defined in the previous section are convenient. The coordinate $\eta$
ranges from $0$ to $\infty$, while both $\xi$ and $\phi$ range from $0$ to
$2\pi$. From the relations
\begin{eqnarray} 
(z-a\textrm{cot}\xi)^2+\rho^2&=&\frac{a^2}{\textrm{sin}^2\xi},\nonumber\\
z^2+(\rho-a\textrm{coth}\eta)^2&=&\frac{a^2}{\textrm{sinh}^2\eta},\nonumber\\
\rho^2&=&x^2+y^2,
\end{eqnarray}
we see that the subspaces defined by $\xi=\textrm{const}$ and $\eta=\textrm{const}$ correspond
to a sphere and a torus, respectively. From the formula
\begin{equation} 
r^2\equiv x^2+y^2+z^2=a^2\frac{\textrm{cosh}\eta+\textrm{cos}\xi}{\textrm{cosh}\eta-\textrm{cos}\xi},
\end{equation}
we understand that $r=0$ and $r=\infty$ correspond to $(\eta=0,\xi=\pi)$
and $(\eta=0,\xi=0)$, respectively.

To define the Hopf index, it is convenient to introduce real fields
$\Phi_{\alpha}(\boldsymbol x)\hspace{1mm} (\alpha=1,2,3,4)$ satisfying
\begin{equation} 
\sum\limits_{\alpha=1}^{4}[\Phi_{\alpha}(\boldsymbol x)]^2=1.
\end{equation}
We next introduce complex fields $Z_1(\boldsymbol x)$, $Z_2(\boldsymbol
x)$ and a column vector $Z(\boldsymbol x)$ by
\begin{eqnarray} 
Z_1(\boldsymbol x)&=&\Phi_1(\boldsymbol x)+i\Phi_2(\boldsymbol
 x),\nonumber\\
Z_2(\boldsymbol x)&=&\Phi_3(\boldsymbol x)+i\Phi_4(\boldsymbol
 x),\nonumber\\
Z(\boldsymbol x)&=&\binom{Z_1(\boldsymbol x)}{Z_2(\boldsymbol x)}.
\end{eqnarray}
The condition (2.3) ensures that the fields $n^a\hspace{1mm}(a=1,2,3)$ defined
by
\begin{equation}
n^a(\boldsymbol x)=Z^{\dagger}({\boldsymbol x})\sigma^a Z({\boldsymbol x}),
\end{equation}
with $\sigma^a\hspace{1mm}(a=1,2,3)$ being Pauli matrices, satisfy the
condition (1.2). More explicitly, ${\boldsymbol n}$ is given by
\begin{equation}
{\boldsymbol n}=\biggl(\frac{u+u^{*}}{|u|^2+1},\frac{-i(u-u^{*})}{{|u|}^2+1},\frac{{|u|}^2-1}{{|u|}^2+1}\biggr),
\end{equation}
where the complex function $u(\boldsymbol x)$ is defined by
\begin{equation}
u(\boldsymbol x)=\biggl({\frac{Z_1(\boldsymbol x)}{Z_2(\boldsymbol x)}}\biggr)^{*}.
\end{equation}
If we define the vector potential ${\boldsymbol A}({\boldsymbol
x})=(A_1({\boldsymbol x}),A_2({\boldsymbol x}), A_3({\boldsymbol x}))$ by
\begin{equation}
A_i({\boldsymbol x})=\frac{1}{i}\{Z^{\dagger}(\boldsymbol x)[\partial_i Z(\boldsymbol x)]-[\partial_i Z^{\dagger}(\boldsymbol x)]Z(\boldsymbol x)\},
\end{equation}
we see that
$H_{ij}(\boldsymbol x)$ defined by 
\begin{equation}
H_{ij}(\boldsymbol x)=\partial_i A_j(\boldsymbol x)-\partial_j A_i(\boldsymbol x)
\end{equation}
coincides with ${\boldsymbol n}\cdot (\partial_i{\boldsymbol
n}\times\partial_j{\boldsymbol n})$ implied by Eq. (1.6). The Hopf index
$Q_H[{\boldsymbol n}]$ is now defined by
\begin{eqnarray}
Q_H[{\boldsymbol n}]&=&\frac{1}{16\pi^2}\int dV {\boldsymbol
 A}({\boldsymbol x})\cdot{\boldsymbol B}({\boldsymbol x}),\\
B_i(\boldsymbol x)&=&\frac{1}{2}\epsilon_{ijk}H_{jk}(\boldsymbol x).
\end{eqnarray}
We now decompose ${\boldsymbol A}({\boldsymbol x})$ as
\begin{equation}
{\boldsymbol A}({\boldsymbol x})={\boldsymbol N}({\boldsymbol x})+{\boldsymbol R}({\boldsymbol x}),
\end{equation}
where ${\boldsymbol N}({\boldsymbol x})$ and ${\boldsymbol
R}({\boldsymbol x})$ satisfy
\begin{eqnarray}
\textrm{rot}{\boldsymbol N}&=&0,\nonumber\\
\textrm{div}{\boldsymbol R}&=&0,
\end{eqnarray}
and
\begin{equation}
\textrm{rot}{\boldsymbol R}=\boldsymbol B.
\end{equation}
Assuming that $|{\boldsymbol x}|^2|{\boldsymbol A}({\boldsymbol x})|$ is
bounded for large $|\boldsymbol x|$, we have
\begin{equation}
{\boldsymbol R}({\boldsymbol x})=\frac{1}{4\pi}\int dV'\frac{1}{|{\boldsymbol x}-{\boldsymbol x'}|}\textrm{rot}{\boldsymbol B}({\boldsymbol x'}),
\end{equation}
and \cite{Kundu}
\begin{eqnarray}
Q_H[\boldsymbol n]&=&\frac{1}{16\pi^2}\int dV{\boldsymbol R}({\boldsymbol x})\cdot {\boldsymbol B}({\boldsymbol x}),\nonumber\\
&=&\frac{1}{64\pi^3}\int dV\int dV'
\frac{1}{|{\boldsymbol x}-{\boldsymbol x'}|^3}({\boldsymbol x}-{\boldsymbol x'})\cdot[{\boldsymbol B}({\boldsymbol x})\times {\boldsymbol B}({\boldsymbol x'})].
\end{eqnarray}
Although there is no local formula expressing ${\boldsymbol
A}(\boldsymbol x)$ in terms of ${\boldsymbol n}(\boldsymbol x)$, the
last formula implies that the Hopf index is calculated solely in terms
of ${\boldsymbol n}$. Another formula for $Q_H[{\boldsymbol n}]$ is \cite{FaddeevAJ}
\begin{equation}
Q_H[\boldsymbol n]=\frac{1}{12 \pi^2}\int dV \epsilon_{\alpha\beta\gamma\delta}\Phi_{\alpha}\frac{\partial{(\Phi_{\beta},\Phi_{\gamma},\Phi_{\delta})}}{\partial(x,y,z)},
\end{equation}
where $\epsilon_{\alpha\beta\gamma\delta}$ is the four-dimensional
Levi-Civita symbol satisfying $\epsilon_{1234}=1$. From this formula, we
can see that the allowed values of $Q_H[\boldsymbol n]$ for regular
$\Phi_\alpha(\boldsymbol x)\hspace{1mm}(\alpha=1,2,3,4)$ are integers.

The AFZ ansatz \cite{AFZ} is stated as follows:
\begin{eqnarray}
Z_1(\boldsymbol x)&=&\frac{f(\eta)}{\sqrt{1+f^2(\eta)}}e^{im\xi},\nonumber\\
Z_2(\boldsymbol x)&=&\frac{1}{\sqrt{1+f^2(\eta)}}e^{-in\phi},
\end{eqnarray}
where $f(\eta)$ is a non-negative function of $\eta$ and $m$ and $n$ are
integers. Then we have
\begin{eqnarray}
u(\boldsymbol x)&=&f(\eta)e^{-i\psi},\nonumber\\
\psi&=&\psi(\xi,\phi)=m\xi+n\phi.
\end{eqnarray}
It is straightforward to obtain
\begin{equation}
dV\epsilon_{\alpha\beta\gamma\delta}\Phi_{\alpha}\frac{\partial{(\Phi_{\beta},\Phi_{\gamma},\Phi_{\delta})}}{\partial(x,y,z)}=3 d\eta\hspace{0.5mm}d\xi\hspace{0.5mm}d\phi\hspace{0.5mm}mn \frac{d}{d\eta}\biggl(\frac{1}{1+f^2(\eta)}\biggr),
\end{equation}
where we have made use of the formula 
$dV=({a}/{q})^3\textrm{sinh}\eta\hspace{0.3mm}
d\eta\hspace{0.3mm} d\xi\hspace{0.3mm}
 d\phi.$
Then we see that the AFZ ansatz with the boundary condition
\begin{equation}
f(0)=\infty,\quad f(\infty)=0
\end{equation}
yields
\begin{eqnarray}
Q_H[\boldsymbol n]&=& mn\biggl(\frac{1}{f^2(\infty)+1}-\frac{1}{f^2(0)+1}\biggr)\nonumber\\
&=& mn.
\end{eqnarray}
As expected, the value of $Q_H[\boldsymbol n]$ is independent of the
detailed structure of $f(\eta)$.

\section{Models compatible with the Aratyn-Ferreira-Zimerman Ansatz}

In the previous section, we saw that the AFZ ansatz associated
with the boundary condition (2.21) automatically
yields configurations ${\boldsymbol n}(\boldsymbol x)$ with the Hopf
index $mn$. Of course, the form of the ansatz for the solution of a field
equation strongly depends on the form of the field equation. In this
section, we show that the AFZ ansatz is applicable to some models other
than that of ${\mathcal L}_{\textrm{AFZ}}(x)$.

We first consider the static field equation derived from the Lagrangian density
\begin{equation}
{\mathcal L}_2^{\kappa}(x)=\biggl(\frac{q}{a}\biggr)^{\kappa}\mu(\eta)\hspace{1mm}l_2(x),
\end{equation}
where $\kappa$ is a real number. Noting that $\epsilon_2$ of Eq. (1.8) is expressed
as
\begin{equation}
\epsilon_2=\frac{4}{(1+|u|^2)^2}(\boldsymbol{\nabla} u\cdot\boldsymbol{\nabla} u^{*}),
\end{equation}
we have the static field equation
\begin{eqnarray}
& &\boldsymbol{\nabla}\cdot(q^{\kappa}j\boldsymbol{\nabla} u^{*})+\frac{1}{2}q^{\kappa}\mu\frac{u^*}{1+|u|^2}\epsilon_2=0,\nonumber\\
& &j=\frac{\mu(\eta)}{(1+|u|^2)^2}.
\end{eqnarray}
When we make use of the toroidal coordinates, the gradient of a function
$g(\boldsymbol x)$ and the divergence of a vector field
\begin{equation} 
{\boldsymbol v}(\boldsymbol x)=v_\eta {\boldsymbol e}_\eta+v_\xi {\boldsymbol e}_\xi+ v_\phi {\boldsymbol e}_\phi
\end{equation}
are given by
\begin{eqnarray}
\boldsymbol{\nabla}g&=&\frac{q}{a}\biggl(\frac{\partial g}{\partial \eta}{\boldsymbol e}_\eta+\frac{\partial g}{\partial \xi}{\boldsymbol e}_\xi+\frac{1}{\textrm{sinh}\eta}\frac{\partial g}{\partial \phi}{\boldsymbol e}_\phi\biggr),\nonumber\\
\boldsymbol{\nabla}\cdot \boldsymbol{v}&=&\frac{q}{a}\biggl[\biggl(\frac{\partial}{\partial \eta}+{\textrm {coth}\eta}-\frac{2\textrm {sinh}\eta}{q}\biggr)v_\eta
+\biggl(\frac{\partial}{\partial \xi}-\frac{2\textrm {sin}\xi}{q}\biggr)v_\xi+\frac{1}{\textrm {sinh}\eta}\frac{\partial v_\phi}{\partial\phi}\biggr],
\end{eqnarray}
where ${\boldsymbol e}_\eta$, ${\boldsymbol e}_\xi$ and ${\boldsymbol
e}_\phi$ are orthonormal unit vectors. If we adopt the AFZ ansatz, the
field equation becomes
\begin{eqnarray}
&&\frac{q^{\kappa+2}}{a^2}e^{i\psi}\biggl(S+i({\kappa}-1)mjf\frac{\textrm{sin}\xi}{q}\biggr)=0,\\
&&S=\biggl(\frac{d}{d\eta}+\textrm{coth}\eta+\frac{({\kappa}-1)\textrm{sinh}\eta}{q}\biggr)(jf')
-Rjf+2jf\frac{{f'}^2+Rf^2}{1+f^2},\\
&&R=m^2+\frac{n^2}{\textrm{sinh}^2\eta},
\end{eqnarray}
where the prime denotes the derivative with respect to $\eta$.
It is clear that, for ${\kappa}\neq 1,m\neq 0$, the unique solution of
Eq. (3.6) is
$f(\eta)\equiv 0$. If and only if ${\kappa}=1$, we have a nontrivial equation
for $f(\eta)$:
\begin{eqnarray}
&&\biggl(\frac{d}{d\eta}+\textrm{coth}\eta\biggr)\biggl(\frac{\mu f'}{(1+f^2)^2}\biggr)+\frac{2\mu f}{(1+f^2)^3}({f'}^2+Rf^2)
-\frac{\mu Rf}{(1+f^2)^2}=0.
\end{eqnarray}
We have thus found that the model defined by
\begin{equation}
\mathcal{L}_2(x)=\mathcal{L}_2^1(x)
\end{equation}
in Eq. (1.16) is compatible with the AFZ ansatz. We note that the static
energy functional
\begin{eqnarray}
E_2[\boldsymbol n]&=&\int dV\hspace{1mm} \frac{q}{a}\mu\hspace{1mm} \epsilon_2\nonumber\\
&=& 16\pi^2\int_0^{\infty}d\eta\hspace{1mm}\mu\hspace{1mm} \textrm{sinh}\eta\hspace{1mm}\frac{{f'}^2+Rf^2}{(1+f^2)^2}
\end{eqnarray}
is independent of the scale parameter $a$.

We next consider the static field equation corresponding to
\begin{equation}
{\mathcal L}_4^{\kappa}(x)=\biggl(\frac{q}{a}\biggr)^{\kappa}\nu(\eta)\hspace{1mm}l_4(x).
\end{equation}
Since $\epsilon_4$ of Eq. (1.9) is given by
\begin{equation}
\epsilon_4=-8\frac{({\boldsymbol \nabla}u\times{\boldsymbol \nabla}u^*)^2}{(1+|u|^2)^4},
\end{equation}
we have
\begin{equation}
{\boldsymbol \nabla}\cdot(q^{\kappa} k{\boldsymbol V})-\frac{1}{4}q^{\kappa}\nu\frac{u^*}{1+|u|^2}\epsilon_4=0,
\end{equation}
where $k$ and $\boldsymbol V$ are defined by
\begin{eqnarray}
k&=&\frac{\nu}{(1+|u|^2)^4},\nonumber\\
\boldsymbol V&=&({\boldsymbol \nabla}u^*)^2{\boldsymbol \nabla}u-({\boldsymbol \nabla}u\cdot{\boldsymbol \nabla}u^*){\boldsymbol \nabla}u^*.
\end{eqnarray}
Under the AFZ ansatz, it becomes
\begin{eqnarray}
& &\frac{q^{{\kappa}+4}}{a^4}e^{i\psi}\biggl(T+i({\kappa}+1)m\frac{\textrm{sin}\xi}{q} k f {f'}^2\biggr)=0,\\
& &T=\biggl(\frac{d}{d\eta}+\textrm{coth}\eta+\frac{{\kappa}+1}{q}\textrm{sinh}\eta\biggr)(k Rf^2f')-kRf {f'}^2 
+ 4 kR\frac{f^3{f'}^2}{1+f^2}.
\end{eqnarray}
In this case, we see that $f(\eta)$ should be trivial $[f(\eta)\equiv
0]$ unless ${\kappa}$ is equal to $-1$. If ${\kappa}=-1$, we obtain the field equation
\begin{equation}
\biggl(\frac{d}{d\eta}+\textrm{coth}\eta\biggr)(k Rf^2f')+kRf\frac{3f^2-1}{f^2+1} {f'}^2=0.
\end{equation}
Thus we see that the model defined by
\begin{equation}
{\mathcal L}_4(x)={\mathcal L}_4^{-1}(x)
\end{equation}
is compatible with the AFZ ansatz. The static energy functional
corresponding to ${\mathcal L}_4(x)$ given by
\begin{eqnarray}
E_4[\boldsymbol n]&=&\int dV \frac{a}{q}\hspace{1mm}\nu \hspace{1mm}\epsilon_4\nonumber\\
&=&128\pi^2\int_0^{\infty} d\eta\hspace{0.5mm}\nu\hspace{0.5mm} R\hspace{0.5mm}\textrm{sinh}\eta\frac{f^2{f'}^2}{(1+f^2)^4}
\end{eqnarray}
is also independent of the scale parameter $a$. It is clear that the AFZ
ansatz is applicable to the model defined by the Lagrangian density
${\mathcal L}_{2+4}(x)$ of  Eq. (1.18).

\section{Solutions of the ${\mathcal L}_2$ models} 

With the aid of the AFZ ansatz, we have obtained the field equation (3.9)
for the model defined by ${\mathcal L}_2(x)$ in Eq. (1.16). The field
equation can also be derived by requiring
\begin{equation}
\delta_f E_2[\boldsymbol n]=0,
\end{equation}
where $E_2[\boldsymbol n]$ is defined by Eq. (3.11) and $\delta_f$ denotes the
variations $f(\eta)\rightarrow f(\eta)+\delta f(\eta)$ and
$f'(\eta)\rightarrow f'(\eta)+\delta f'(\eta)$ with other factors
fixed. We here define a new variable by
\begin{eqnarray}
t(\eta)&=&2|m|\textrm{log} z+|n|\textrm{log}\biggl(\frac{z^2-A^2}{z^2-B^2}\biggr),\nonumber\\
z&=&\textrm{cosh}\eta+\sqrt{\textrm{sinh}^2 \eta+\frac{n^2}{m^2}},\nonumber\\
A&=&\frac{|n|+|m|}{|m|},\quad B=\frac{|n|-|m|}{|m|}.
\end{eqnarray}
In the case that $|m|=|n|$, $t(\eta)$ simplifies to $2|m|$$\textrm{log}(2\textrm{sinh}\eta)$.
 The variable $t(\eta)$ is a solution of the equation
\begin{equation}
\frac{dt(\eta)}{d\eta}=2\sqrt{R(\eta)},
\end{equation} 
where $R(\eta)$ is defined by Eq. (3.8). In terms of $t(\eta)$ and $w(t)$ defined by
\begin{equation}
w(t)=-f^2(\eta),
\end{equation} 
Eq. (3.9) is rewritten as
\begin{eqnarray}
\frac{d^2w}{dt^2}&=&\biggl(\frac{1}{2w}+\frac{1}{w-1}\biggr)
\biggl(\frac{dw}{dt}\biggr)^2-\frac{dm}{dt}\frac{dw}{dt}
-\frac{1}{2}
\frac{w(w+1)}{w-1},\\
m(t)&=&\textrm{log}\biggl(\mu(\eta)\sqrt{R(\eta)}\textrm{sinh}\eta\biggr).
\end{eqnarray}
Various ways of choosing $\mu(\eta)$ lead us to some interesting cases.
 We here describe two of them.

\subsection{Case solved by the fifth Painlev\'e transcendent}

In a celebrated paper of 1976, Wu, McCoy, Tracy and Barouch \cite{Wu} discovered that the scaling
limit of the correlation function of the two-dimensional Ising model is described by a solution of 
the 
$(\alpha,\beta,\gamma,\delta)=(0,0,1,-1)$ case of the third Painlev\'e
equation 
$\textrm{P}_{\textrm{III}}\hspace{0.5mm}(\alpha,\beta,\gamma,\delta)$ \cite{Ince,Iwasaki}:
\begin{equation}
\frac{d^2 y}{dx^2}=\frac{1}{y}\biggl(\frac{dy}{dx}\biggr)^2-\frac{1}{x}\frac{dy}{dx}+\frac{1}{x}(\alpha y^2+\beta)+\gamma y^3+\frac{\delta}{y},
\end{equation}
where $\alpha$, $\beta$, $\gamma$, and $\delta$ are parameters. Since
then, six types of Painlev\'e equation have appeared here and there in
mathematical physics \cite{Sato,Dubrovin}.

If we set
\begin{equation}
\mu(\eta)=\textrm{const}\frac{\textrm{log}\hspace{0.5mm}(2\textrm{sinh}\eta)}{\textrm{cosh}\eta},
\end{equation}
then Eq. (4.5) for the case $|m|=|n|$ becomes
\begin{equation}
\frac{d^2w}{dt^2}=\biggl(\frac{1}{2w}+\frac{1}{w-1}\biggr)
\biggl(\frac{dw}{dt}\biggr)^2
-\frac{1}{t}\frac{dw}{dt}-\frac{1}{2}\frac{w(w+1)}{w-1}.
\end{equation}
This equation is nothing but the
 $(\alpha,\beta,\gamma,\delta)=(0,0,0,$
$-\frac{1}{2})$
case of the fifth Painlev\'e equation
 $\textrm{P}_{\textrm{V}}(\alpha,\beta,\gamma,\delta)$:
\begin{eqnarray}
\frac{d^2y}{dx^2}&=&\biggl(\frac{1}{2y}+\frac{1}{y-1}\biggr)
\biggl(\frac{dy}{dx}\biggr)^2
-\frac{1}{x}\frac{dy}{dx}+\frac{(y-1)^2}{x^2}
\biggl(\alpha y+\frac{\beta}{y}\biggr)
+
\frac{\gamma y}{x}+\delta\frac{y(y+1)}{y-1}.
\end{eqnarray}
$\textrm{P}_{\textrm{V}}(\alpha,\beta,\gamma,\delta)$ with $\delta\neq 0$ cannot be reduced to simpler equations
 and, 
without loss of generality, $\delta$ can be assumed to be $-\frac{1}{2}$.

It is well known that the solutions of the Painlev\'e equations have
movable poles, that is, 
poles whose positions depend on the boundary or initial conditions
imposed. If we assume that
$w(t)$ has a zero at $t_0\neq 0,\infty$ and a pole at $t_1\neq 0,\infty$, we
find that both of them must be of second order:
\begin{eqnarray}
&&w(t)\sim \textrm{const}\times(t-t_0)^2 \hspace{8mm} (t\sim t_0),\nonumber\\
&&w(t)\sim \textrm{const}\times\frac{1}{(t-t_1)^2}\hspace{7mm} (t\sim t_1).
\end{eqnarray}
Near the pole $t_1$, it is convenient to define
\begin{equation}
v(t)=\frac{1}{w(t)}.
\end{equation}
It is interesting that Eq. (4.9) is converted to an equation of the same form:
\begin{equation}
\frac{d^2 v}{dt^2}=\biggl(\frac{1}{2v}+\frac{1}{v-1}\biggr)
\biggl(\frac{dv}{dt}\biggr)^2
-\frac{1}{t}\frac{dv}{dt}-\frac{1}{2}\frac{v(v+1)}{v-1}.
\end{equation}
We note that Eq. (4.9) or Eq. (4.13) is obtained by eliminating $\sigma$ from
the Hamiltonian system \cite{Okamoto}
\begin{eqnarray}
\frac{d\lambda}{dt}&=&\frac{\partial H}{\partial \sigma}\nonumber\\
&=&\frac{\lambda}{t}[2(\lambda-1)
^2\sigma +\lambda-1+t],\nonumber\\
\frac{d\sigma}{dt}&=&-\frac{\partial H}{\partial \lambda}\nonumber\\
&=&-\frac{1}{t}\biggl[(\lambda-1)
(3\lambda-1)\sigma^2 
+(2\lambda+t-1)\sigma+\frac{1}{4}\biggr]
\end{eqnarray}
with 
\begin{eqnarray}
H&=&H(\lambda,\sigma,t)\nonumber\\
&=&\frac{1}{t}\biggl\{\lambda(\lambda-1)^2\sigma^2+[\lambda(\lambda-1)+t\lambda
]\sigma
+\frac{1}{4}(\lambda-1)\biggr\}.
\end{eqnarray}
We here describe an example of numerical analysis aided by a personal computer.
If we set the initial condition as
\begin{equation}
w=-10^{-6},\quad \frac{dw}{dt}=-0.002 \hspace{3mm}\textrm{at}\hspace{3mm}t=10,
\end{equation}
that is,
\begin{equation}
v=-10^{6},\quad \frac{dv}{dt}=2\times 10^9 \hspace{3mm}\textrm{at}\hspace{3mm}t=10,
\end{equation}
we find that $|w(t)|$ becomes very close to zero at
\begin{equation}
t_0\fallingdotseq 9.99900, 
\end{equation}
and blows up at
\begin{equation}
t_1\fallingdotseq 11.6723.
\end{equation}
Then we can think of the following configuration of ${\boldsymbol n}(\boldsymbol x)
$:
\begin{eqnarray}
{\boldsymbol n}(\boldsymbol x)&=&(0,0,-1)\quad;\quad \eta\leqq \eta_0,\nonumber\\
{\boldsymbol n}(\boldsymbol x)&=&\biggl(\frac{2\sqrt{-w(t)}}{1-w(t)}\textrm{cos}(m\xi+n\phi),
\frac{-2\sqrt{-w(t)}}{1-w(t)}
\textrm{sin}(m\xi+n\phi),\frac{w(t)+1}{w(t)-1}\biggr);\nonumber\\
&&\quad \eta_0<\eta<\eta_1,\nonumber\\
{\boldsymbol n}(\boldsymbol x)&=&(0,0,1);\quad \eta_1\leqq \eta,
\end{eqnarray}
where $\eta_i\hspace{1mm}(i=0,1)$ is defined by $t(\eta_i)=t_i$ and $n$
is equal to $m$ or $-m$. The energy carried by the above configuration is given by
\begin{eqnarray}
E_{2A}[\boldsymbol n]&=&8\pi^2\int_{t_0}^{t_1} dt\hspace{1mm} \frac{t}{|w(t)|[1+|w(t)|]^2}
\{[{\dot w(t)}]^2+[w(t)]^2\},
\end{eqnarray}
where the overdot denotes the derivative with respect to $t$. The behavior (4.11) of $w(t)$ near $t_0$ and $t_1$ ensures that $E_2[\boldsymbol n]$ is finite. We have thus seen that the above configuration corresponds to a soliton in which the energy distribution is strictly confined within the finite spatial volume
\begin{equation}
V=\{ (\eta,\xi,\phi)|\hspace{1mm}\eta_0<\eta<\eta_1,\hspace{1mm}0\leqq 
\xi,\phi
\leqq 2\pi\}.
\end{equation}
To be more precise, although ${\boldsymbol n}(\boldsymbol x)$ in Eq. (4.20) is continuous in the whole ${\boldsymbol R}^3$, its gradient ${\boldsymbol \nabla}{\boldsymbol n}(\boldsymbol 
x)$ is discontinuous on the surfaces $\eta=\eta_0$ and
$\eta=\eta_1$. Therefore we have to modify $\mu(\eta)$ in the
neighborhood of these surfaces so as to make ${\boldsymbol
\nabla}{\boldsymbol n}(\boldsymbol x)$ continuous. The above
neighborhood can, however, be made as thin as we want. Then a
discussion similar to that around Eq. (2.22) would yield the result
$Q_H[\boldsymbol n]=-mn$. It should be noted that, for the choice of
Eq. (4.8), the large-$\eta$ behavior of $m(t)$ for general $|n/m|\neq
0,\infty$ is given by $m(t)\sim \textrm{log}\hspace{0.5mm}t$ and we have
Eq. (4.9).

\subsection{Case solved by the Jacobi elliptic function}

If we set $\mu(\eta)$ as
\begin{equation}
\mu(\eta)=\frac{\textrm{const}}{\textrm{cosh}\eta},
\end{equation}
then Eq. (4.5) for the case of $|m|=|n|$ becomes
\begin{equation}
\frac{d^2w}{dt^2}=\biggl(\frac{1}{2w}+\frac{1}{w-1}\biggr)\biggl(\frac{dw}{dt}
\biggr)^2-\frac{1}{2}\frac{w(w+1)}{w-1}.
\end{equation}
This equation can be derived by extremizing
\begin{eqnarray}
E_{2B}[{\boldsymbol n}]&=&8\pi^2\int_{-\infty}^{\infty} dt\hspace{1mm} l\bigl(w(t),\dot{w}(t)\bigr),\\
l(w,\dot{w})&=&-\frac{1}{w(1-w)^2}({\dot w}^2+w^2).
\end{eqnarray}
Then, defining $p(t)$ and $h\bigl(w(t),p(t)\bigr)$ by
\begin{eqnarray}
p&=&\frac{\partial l(w,\dot w)}{\partial{\dot w}}=-\frac{2\dot w}{w(1-w)^2},\\
h(w,p)&=&p\dot w-l(w,\dot w)\nonumber\\
&=&-\frac{1}{4}w(1-w)^2p^2+\frac{w}{(1-w)^2},
\end{eqnarray}
we have
\begin{equation}
h(w,p)=\textrm{const}\equiv -c.
\end{equation}
From Eqs. (4.27), (4.28), and (4.29), we obtain
\begin{equation}
{\dot w}^2=w^2+cw(1-w)^2,
\end{equation}
which is solvable in terms of the Jacobi elliptic function
$\textrm{sn}(x,k)$. We first rewrite Eq. (4.30) as
\begin{equation}
{\dot w}^2=cw(w-a)(w-b),
\end{equation}
where $a$ and $b$ are defined by
\begin{equation}
a=1-\frac{1}{2c}-\sqrt{\frac{1-4c}{4c^2}}=\frac{1}{b}.
\end{equation}
When $c$ satisfies $\frac{1}{4}>c>0$, we readily find that 
\begin{eqnarray}
w(t)&=&\frac{1}{b}-\frac{1-b^2}{b}\textrm{sn}^2\biggl(\frac{1}{2}\sqrt{-\frac{c}{b}}\hspace{1mm}t,\sqrt{1-b^2}\biggr),\nonumber\\
&&-1<b<0,
\end{eqnarray}
is a solution. In the case of $c=0$, we see that
\begin{equation}
w_{-}(t)=-de^{-t},\hspace{2mm}d>0,
\end{equation}
and 
\begin{equation}
w_{+}(t)=-de^{t},\hspace{2mm}
\end{equation}
are solutions. In the case that $c<0$, we find that
\begin{equation}
w(t)=-\frac{1}{b}
\frac{\textrm{sn}^2\biggl[({\sqrt{-bc}}/{2})t,{\sqrt{b^2-1}}/{b}\biggr]}
{1-\textrm{sn}^2\biggl[({\sqrt{-bc}}/{2})t,{\sqrt{b^2-1}}/{b}\biggr]},\hspace{2mm}b>1,
\end{equation}
is a solution. We see, however, that there is no nontrivial solution
for $c\geqq \frac{1}{4}$.

\section{Solutions of the ${\mathcal L}_4$ models} 

The field equation of the model defined by $\mathcal{L}_4(x)$ was given
by Eq. (3.18). It is not difficult to rewrite this equation as
\begin{equation}
\frac{f''}{f'}+\frac{(1-3f^2)f'}{(1+f^2)f}+\frac{r'}{r}=0,
\end{equation}
where $r=r(\eta)$ is defined by
\begin{equation}
r(\eta)=R(\eta)\nu(\eta)\textrm{sinh}\eta.
\end{equation}
Integrating Eq. (5.1), we have
\begin{equation}
\frac{rff'}{(f^2+1)^2}=\textrm{const}\equiv c_1.
\end{equation}
If we set the boundary condition for $f(\eta)$ as $f(\infty)=0$, $f(0)=\infty$, we obtain the solution
\begin{equation}
f^2(\eta)=\frac{\int_{\eta}^\infty d\eta\hspace{1mm}s(\eta)}{\int_0^\eta 
d\eta\hspace{1mm}s(\eta)},
\end{equation}
where $s(\eta)$ is defined by
\begin{equation}
s(\eta)=\frac{1}{r(\eta)}=\frac{\textrm{sinh}\eta}{\nu(\eta)(m^2\textrm{sinh}^2\eta+n^2)}.
\end{equation}
As was stated in Sec. II, the Hopf index $Q_H[\boldsymbol n]$ is equal to $mn$ in this case.
 We have thus found an infinite number of solutions with arbitrary values of 
$Q_H[\boldsymbol n]$. From Eqs. (5.3) and (5.4), we have
\begin{equation}
c_1=-\frac{1}{2\int_0^\infty d\eta\hspace{1mm} s(\eta)}.
\end{equation}
Then the energy functional $E_4[\boldsymbol n]$ in Eq. (3.20) is given by
\begin{equation}
E_4[\boldsymbol n]=\frac{32\pi^2}{\int_0^\infty d\eta\hspace{1mm}s(\eta)}.
\end{equation}
It is interesting that the above expression does not involve the function $f(\eta)$.
We see that $f(\eta)$ and $E_4[\boldsymbol n]/m^2$ are determined by $\nu(\eta)$
 and 
$|m/n|$. In the simplest case with $\nu(\eta)=1$ and $|m/n|=1$, we have
\begin{equation}
f^2(\eta)=\frac{1}{\textrm{cosh}\eta-1}
\end{equation}
and 
\begin{equation}
E_4[\boldsymbol n]=32\pi^2m^2.
\end{equation}

\section{$\mathcal{L}_{2+4}$ case}

We have seen that the $\mathcal{L}_2(x)$ model is solvable for some
choices of $\mu(\eta)$ and that the $\mathcal{L}_4(x)$ model is solvable
for general $\nu(\eta)$. Here we consider the $\mathcal{L}_{2+4}(x)$
model defined by Eq. (1.18). The static energy functional
$E_{2+4}[\boldsymbol{n}]$ corresponding to $\mathcal{L}_{2+4}(x)$ is
given by
\begin{eqnarray}
E_{2+4}[\boldsymbol{n}] &=&g_2E_2[\boldsymbol{n}] +g_4E_4[\boldsymbol{n}]\nonumber\\
&=& 16\pi^2\int_0^\infty d\eta \hspace{1mm}\textrm{sinh}\eta\biggl(g_2\mu\frac{{f'}^2+Rf^2}{(1+f^2)^2}
+8g_4\nu\frac{Rf^2{f'}^2}{(1+f^2)^4}\biggr). 
\end{eqnarray}
In terms of $t(\eta)$ and $w(t)$ defined in the previous section, $E_{2+4}[\boldsymbol{n}]$ is expressed as
\begin{equation}
E_{2+4}[\boldsymbol{n}] =8\pi^2g_2\int_{-\infty}^\infty dt\hspace{1mm} l_{2+4}(w,\dot{w},t),
\end{equation}
where $l_{2+4}(w,\dot{w},t)$ is given by
\begin{eqnarray}
l_{2+4}(w,\dot{w},t)&=&\lambda\biggl(-\frac{{\dot w}^2+w^2}{w(1-w)^2}+\sigma\frac{{\dot w}^2}{(1-w)^4}\biggr),\nonumber\\
\lambda&=&\lambda(t)=\mu(\eta)\sqrt{R(\eta)}\hspace{1mm}\textrm{sinh}\eta,\nonumber\\
\sigma&=&\sigma(t)=\frac{8g_4}{g_2}\frac{\nu(\eta)R(\eta)}{\mu(\eta)}.
\end{eqnarray}
As in the discussion of the previous sections, the field equation is given by
\begin{equation}
\frac{d}{dt}\biggl(\frac{\partial{l_{2+4}}(w,{\dot w},t)}{\partial {\dot w}}\biggr)-\frac{\partial{l_{2+4}}(w,\dot w,t)}{\partial w}=0
\end{equation}
or
\begin{eqnarray}
\frac{dw}{dt}&=&\frac{\partial h_{2+4}(w,p,t)}{\partial p},\nonumber\\
\frac{dp}{dt}&=&-\frac{\partial h_{2+4}(w,p,t)}{\partial w}
\end{eqnarray}
with $p$ and $h_{2+4}$ being defined by
\begin{eqnarray}
p&=&\frac{\partial {l_{2+4}}(w,{\dot w},t)}{\partial {\dot w}},\nonumber\\
h_{2+4}(w,p,t)&=&{\dot w}p-l_{2+4}(w,{\dot w},t).
\end{eqnarray}
The resulting equation is given by
\begin{equation}
{\ddot w}+\frac{1}{2g}\frac{\partial g}{\partial w}{\dot w}^2+\biggl(\frac{\dot \lambda}{\lambda}
+\frac{1}{g}\frac{\partial g}{\partial \sigma}\dot\sigma\biggr)\dot w+\frac{1}{2g}\frac{w+1}{(w-1)^3}=0,
\end{equation}
where $g$ is defined by
\begin{equation}
g=g(w,\sigma)=\frac{(w-1)^2-\sigma w}{w(w-1)^4}.
\end{equation} 
The above equation becomes simplest in the case that $\lambda$ and
$\sigma$ are constant. If we set
\begin{eqnarray}
\mu(\eta)&=&\textrm{const}\times\frac{1}{\textrm{cosh}\eta},\nonumber\\
\nu(\eta)&=&\textrm{const}\times\frac{\textrm{sinh}^2\eta}{\textrm{cosh}^3\eta},
\end{eqnarray}
and consider the $|m|=|n|$ case, $l_{2+4}$ and $h_{2+4}$ do not involve $t$ explicitly. Then we have
\begin{eqnarray}
p&=&-2\lambda \biggl(\frac{(1-w)^2-\sigma w}{w(1-w)^4}\biggr)\dot{w},\nonumber\\
h_{2+4}(w,p)&=&-\frac{w(1-w)^4}{4\lambda[(1-w)^2-\sigma w]}\hspace{1mm}p^2
+\lambda\frac{w}{(1-w)^2}=\textrm{const}\equiv c_2,\nonumber\\
&&\hspace{-15mm}\lambda,\sigma=\textrm{const}
\end{eqnarray}
We now obtain
\begin{equation}
{\dot w}^2=-\frac{1}{\lambda} w(w-1)^2\frac{c_2(w-1)^2-\lambda w}{(w-1)^2-\sigma w}\equiv J(w).
\end{equation}
The above equation cannot be solved analytically for general values of
$c_2$. It can be solved, however, for some special values of $c_2$ such
as $0$ and $\lambda/\sigma$. We here assume that  $\lambda$ and $\sigma$ are
positive constants. If the parameter $c_2$ satisfies $-{\lambda}/{4}<c_2<0$, we find
that there exists an oscillating solution  $w(t)\in[{\alpha}^{-1},\alpha]$, where $\alpha$ is defined by
\begin{equation}
\alpha=1+\frac{\lambda}{2c_2}+\sqrt{\frac{\lambda}{c_2}+\frac{\lambda^2}{4{c_2}^2}}
\end{equation}
and satisfies $-1<\alpha<0.$ The period $T$ of the oscillation is given by
\begin{equation}
T=2\int_{{\alpha}^{-1}}^{\alpha}\frac{dw}{\sqrt{J(w)}}.
\end{equation}
On the other hand, when $c_2$ is positive, we have solutions $w_{+}(t)$
and $w_{-}(t)$ determined by
\begin{equation}
\pm\int_{w_{\pm}(t)}^0\frac{dw}{\sqrt{J(w)}}=t_0-t
\end{equation}
and satisfying $w_{\pm}(t_0)=0$ and $w_{\pm}(t_1)=-\infty$, where $t_1$ is
defined by
\begin{equation}
t_1=t_0\mp\int_{-\infty}^0\frac{dw}{\sqrt{J(w)}}.
\end{equation}
These solutions should be compared with those discussed in Sec. IV A. For other values of $c_2$, we have no nontrivial solution. 

\section{summary}

We have investigated some classes of nonlinear $\sigma$ models under
the ansatz proposed by Aratyn, Ferreira, and Zimerman. The models defined
by ${\mathcal L}_2(x)$, ${\mathcal L}_4(x)$, and ${\mathcal L}_{2+4}(x)$
were shown to be compatible with the ansatz. The ${\mathcal L}_2(x)$
model with $\mu(\eta)$ given by Eq. (4.8) was shown to result in the
$(\alpha,\beta,\gamma,\delta)=(0,0,0,-\frac{1}{2})$ case of the fifth
Poinlev\'e equation. The ${\mathcal L}_2(x)$ model with $\mu(\eta)$
given by Eq. (4.23) was solved in terms of the Jacobi elliptic function. The
${\mathcal L}_4(x)$ model was solved for general $\nu(\eta)$. The
${\mathcal L}_{2+4}(x)$ model with $\mu(\eta)$ and $\nu(\eta)$ given by
Eq. (6.9) was shown to have an oscillating solution with the period (6.13)
and monotonic solutions determined by Eq. (6.14). We discussed in Secs. IV
and VI the choices of $\mu(\eta)$ and $\nu(\eta)$ that make
discussion of the solutions with $|m|=|n|$ simple. It is possible,
however, to make choices of $\mu(\eta)$ and $\nu(\eta)$ so that the
discussion of the case $|m/n|\neq 1$ is simple. For example, if we set
$\mu(\eta)=\{t(\eta)/[\sqrt{R(\eta)}\textrm{sinh}\eta]\}_{m=m_0,n=n_0}$ with
$t(\eta)$ and $R(\eta)$ given by Eqs. (4.2) and (3.8), respectively, the
$|m/n|=|m_0/n_0|$ cases of Eq. (4.5) can be written as Eq. (4.9). Similarly, the
$|m/n|=|m_0/n_0|$ case of Eq. (4.5) can be written as Eq. (4.24) if we set $\mu(\eta)=\{1/[\sqrt{R(\eta)}\textrm{sinh}\eta]\}_{m=m_0,n=n_0}$.

\begin{acknowledgments}

The authors are grateful to Shinji Hamamoto, Takeshi Kurimoto, Hiroshi
 Kakuhata, Hitoshi Yamakoshi, Hideaki Hayakawa, and Jun Yamashita for discussions.
\end{acknowledgments} 



\end{document}